\documentclass[mathleft
]{an}
\usepackage{graphicx}
\usepackage{natbib}
\bibpunct{(}{)}{,}{a}{}{,}
\usepackage{times}
\begin{document}

\Pagespan{0}{}
\Yearpublication{}%
\Yearsubmission{}%
\Month{}%
\Volume{}%
\Issue{}%

\title{Observations and interpretation of solar-like oscillations in red-giant stars}

\author{S. Hekker}

\titlerunning{Observations and interpretation of solar-like oscillations in red-giant stars}
\authorrunning{S. Hekker}
\institute{University of Birmingham, School of Physics and Astronomy, Edgbaston Birmingham B15 2TT, United Kingdom (saskia@bison.ph.bham.ac.uk)}

\keywords{stars: oscillations - stars: late type}

\abstract{Over the past decade the study of solar-like oscillations in red-giant stars has developed significantly. Not only the number of red-giant stars for which solar-like oscillations have been observed has increased, but the quality of these observations has improved as well. These steps forward were possible thanks to the development of instrumentation to measure radial velocity variations with a precision of the order of m/s, as well as the launch of dedicated space missions, which provide timeseries of data with unprecedented photometric precision. Many more exciting discoveries are to be expected in the (near) future. This article provides an overview of the development of the field over the last decade, discusses difficulties encountered and overcome in interpreting the observational data, and addresses some challenges and opportunities for further research.}

\maketitle

\section{Introduction}
All stars with masses between roughly 0.6 and 6 solar masses evolve through a red-giant phase. 
Stars in this evolutionary phase follow a relatively narrow track through the H-R diagram, i.e., during the ascent and descent of the giant branch and their time on the horizontal branch, while their internal structures are different as a function of e.g., mass, metallicity and primordial rotational velocity.
The large number of observable red giants, and the drastic changes in their internal structures on relatively fast timescales make these stars ideal for studying stellar evolution. The only way to unravel their internal structures is through study of stellar oscillations. 

For years however, late-G and early-K red giants had been considered to be pulsationaly stable, see e.g. Eyer \& Grenon (1997), Jorissen et al. (1997). Therefore these stars were selected as candidates to act as reference stars for space missions, such as SIM (Space Interferometry Mission (SIM PlanetQuest), \citet{frink2001}). Indeed stable stars with long term (months to years) radial velocity variations of the order of 20 m\,s$^{-1}$ or less were discovered during the SIM preparation survey \citep{hekker2006a}. On a much shorter time scale of the order of hours to days most of these stars do however show pulsational variability. This variability originates from stochastically excited p-mode oscillations in the turbulent outer layers of the stars. Through the study of these, so-called solar-like oscillations, the internal structure of these stars can be investigated. However, until about ten years ago, the quality of observational data was insufficient to perform such study on many giants.
Only for significantly evolved giants, which have oscillations with large amplitudes and relatively long periods of the order of days, such as Arcturus, oscillations could be detected. See for a review of these results \citet{merline1999}.

The oscillations in less evolved red giants became detectable with the development of high-precision instrumentation and oscillations in $\alpha$ Uma (K0III) from photometry with the WIRE satellite \citep{buzasi2000}  and in $\xi$ Hydrae (G7III) from radial velocity measurements \citep{frandsen2002} are among the earlier results mentioned. These observations were followed by theoretical computations by Dziembowski et al. (2001) for $\alpha$ Uma and, by among others, Houdek \& Gough (2002), \citet{teixeira2003} for $\xi$ Hydrae to interpret these observations.

These early observations of solar-like oscillations in red giants initiated radial velocity multi-site campaigns on similar targets \citep{barban2004,deridder2006}, observations with non-dedicated space instruments, such as the star tracker on the WIRE satellite \citep{buzasi2000} and the Solar Mass Ejection Imager (SMEI) on board the Coriolis spacecraft, and observations with dedicated satellites such as MOST \citep{matthews2000}, CoRoT  Baglin et al. (2006) and recently Kepler \citep{borucki2009}. Results of these observations are presented by e.g., \citet{tarrant2007} using SMEI data, \citet{barban2007,kallinger2008} using MOST observations, \citet{deridder2009,hekker2009} using CoRoT data, and \citet{bedding2010,hekker2010a} using Kepler data. 

The interpretation of the observations of solar-like oscillations in red-giant stars from early observations and theoretical computations appeared to be ambiguous. Different interpretations resulted for instance in differences in mode lifetimes of $\xi$ Hydrae of an order of magnitude. This was later followed by a related discussion on the presence and observability of non-radial modes.

\section{Solar-like oscillations}
The oscillations in red giants originate from the same mechanism as the oscillations observed for the Sun, i.e. they are stochastically excited in the turbulent outer layer of the star. In many stars the observed oscillations follow a regular pattern and it seems to be justified to use the asymptotic approximation developed by \citet{tassoul1980} for high-order low-degree modes:
\begin{equation}
\nu_{n,\ell} \approx \Delta \nu \left (n+\frac{1}{2}\ell+\epsilon \right )-\ell(\ell+1)D_0,
\label{asymptot}
\end{equation}
with $\nu$ the oscillation frequency, $n$ the radial order, $\ell$ the angular degree, $\Delta \nu$ the large separation between modes of same degree and consecutive orders, which is inversely proportional to the sound travel time through the star, i.e. a proxy for the average density of the star. $\epsilon$ is sensitive to the surface layers and $D_0$ to deeper layers in the star. $D_0$ can be measured from any of the three small separations which are defined as: $\delta \nu_{02}$ the spacing between adjacent modes with $\ell$~=~0 and $\ell$~=~2, $\delta \nu_{13}$ the spacing between adjacent modes with $\ell$~=~1 and $\ell$~=~3, and $\delta \nu_{01}$ the amount by which  the $\ell$~=~1 modes are offset from the midpoints between the $\ell$~=~0 mode on either side. If Eq.~\ref{asymptot} holds than $\delta \nu_{02}$ = 6$D_0$, $\delta \nu_{13}$ = 10$D_0$ and $\delta \nu_{01}$ = 2$D_0$ \citep{tassoul1980}, in which case the $\ell$~=~1 modes are located on the left side of the midpoint between adjacent $\ell$~=~0 modes. However there is evidence that for red giants  the relations between the three small separations and $D_0$ do not hold, as negative values for $\delta \nu_{01}$ have been observed \citep{carrier2010,bedding2010}. In these cases the $\ell$ = 1 modes are located on the right side of the midpoints between the adjacent $\ell$~=~0 modes. A negative $\delta \nu_{01}$ is most likely a trace of stellar evolution (see Montalban et al. 2010, these proceedings).

When the oscillation frequencies follow a regular pattern as predicted by Eq.~\ref{asymptot}, a so-called \'{e}chelle diagram (Grec et al. 1983)
can be used to investigate the degree of the mode. In an \'{e}chelle diagram the frequencies are represented as a function of the frequency modulo $\Delta \nu$. Oscillation modes of the same degree are then visible in vertical ridges, with the modes with $\ell$~=~0 and $\ell$~=~1 separated by approximately 0.5$\cdot \Delta \nu$ and the $\ell$~=~2 modes close to the $\ell$~=~0 mode (see for an example the bottom panel of Fig.~\ref{deridder}).

\section{Mode lifetimes}
Solar-like oscillations are stochastically excited and have finite mode lifetimes, i.e., the modes are damped, with a rate $e^{-\eta t}$, with $\eta$ the damping rate, and re-excited before the oscillation mode is damped out. Due to this effect both the amplitude and phase of the oscillation are time dependent and the frequency peaks in the power spectrum have Lorentzian shapes with a width inversely proportional to the mode lifetime ($\tau$ $\sim$ $\eta^{-1}$). Therefore, the mode lifetime provides information on the damping processes.

\begin{figure}
\begin{minipage}{\linewidth}
\centering
\includegraphics[width=\linewidth]{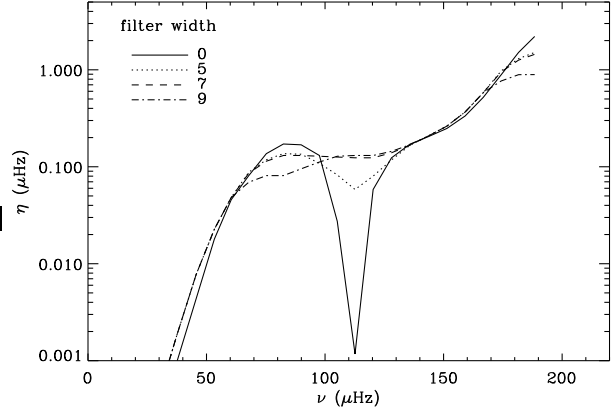}
\end{minipage}
\caption{Theoretical damping rate $\eta$ as a function of frequency for $\xi$ Hydrae. The solid line shows the raw damping rates while the other lines show smoothed rates over the number of radial orders $n$ as indicated in the legend (Figure taken from \citet{houdek2002}). The large dip in the raw damping rates is related to the properties of the superadiabatic boundary layer.}
\label{houdek}
\end{figure}

\citet{houdek2002} modelled the excitation and damping of the stochastic modes observed in $\xi$ Hydrae (Frandsen et al. 2002). From the damping rates (see Fig.~\ref{houdek}) obtained from their models, \citet{houdek2002} computed the mode lifetimes of the observed oscillations. They obtained a mode lifetime of the order of 20 days.

\begin{figure}
\begin{minipage}{\linewidth}
\centering
\includegraphics[width=\linewidth]{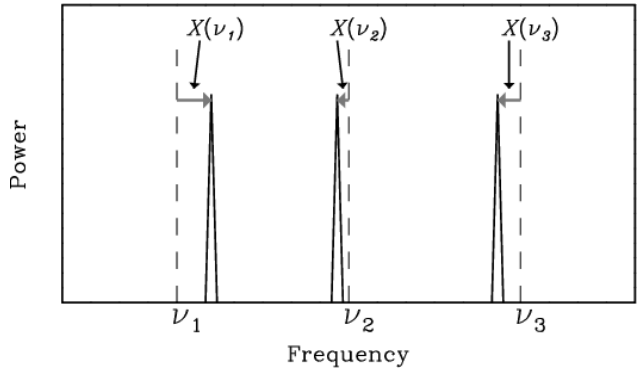}
\end{minipage}
\caption{Schematic illustration of the frequency scatter due to a finite mode lifetime. Dashed lines are the predicted mode frequencies and the solid peaks are the measured frequencies (Figure taken from \citet{stello2006}).}
\label{stello}
\end{figure}

For the same star, \citet{stello2006} developed a new method to measure the mode lifetimes directly from the width of the oscillation features in the power spectrum. For stars with short mode lifetimes the peaks in the power spectrum are wider than for stars with long mode lifetimes. By measuring the scatter of the observed frequencies around the expected regular pattern (Eq.~\ref{asymptot} and Fig.~\ref{stello}) and comparing this scatter with extensive simulations they found that the mode lifetime for $\xi$ Hydrae had to be of the order of 2 days. These results were confirmed by \citet{brewer2009} using Gaussian process modelling.

Observations with the CoRoT satellite have revealed that some red giants have long mode lifetimes ($>$ 50 days) (Baudin et al. 2010, De Ridder et al. 2009), while other red giants have shorter mode lifetimes of the order 15 days \citep{carrier2010}. Although the mode lifetimes of $\xi$ Hydrae could not be verified, these space-based photometric data did show that red giants can have mode lifetimes from the order of 15 days up to a few tens of days or even longer, possibly depending on their evolutionary state.

\section{Non-radial modes}
The Brunt-V\"ais\"al\"a ($N$) and Lamp ($L_{\ell}$) frequencies are huge in the dense core of red giants. Because of these huge values, all non-radial modes with frequencies lower than the cut-off frequency and larger than the fundamental radial mode have mixed character. In the envelope the oscillations behave like acoustic modes, while in the core the same modes behave like gravity modes. The evanescent region between the p- and g-mode cavities is at the origin of mode trapping. Some modes have significant kinetic energy in the g-cavity and low ones in the p-cavity; they are trapped in the core. Others have low kinetic energy in the g-cavity, but high ones in the envelope; they are trapped in the envelope \citet{dupret2009}. 

The theoretical computations of oscillations of $\alpha$ Uma and other red giants by \citet{dziembowski2001} have been the standard theory work for red-giant asteroseismology for nearly a decade. In this work they predict that only radial modes will be observable on the surface of red-giant stars due to the trapping of non-radial oscillation modes in the core of the star. Consequently, the observed oscillation modes of $\xi$ Hydrae \citep{frandsen2002} and $\epsilon$ Ophiuchi (De Ridder et al. 2006) were interpreted as radial oscillations. Hekker et al. (2006a) were the first to investigate the nature of the modes using line profile analysis of the spectra of $\xi$ Hydrae, $\epsilon$ Ophiuchi, $\eta$ Serpentis and $\delta$ Eridani. They concluded that non-radial oscillations were present, although a definite mode identification could not be made. An updated version of this analysis has recently been published \citep{hekker2010b}, which supports the general conclusions of Hekker et al. (2006a).

The analysis of the data taken during the spectroscopic multi-site campaign presented by \citet{deridder2006} revealed two possible models for $\epsilon$ Ophiuchi. Therefore this star was also observed for three weeks with the MOST satellite. These results were interpreted as radial modes with mode lifetimes of the order of a few days \citep{barban2007} and could narrow the solution for $\epsilon$ Ophiuchi to one specific model. Subsequently, \citet{kallinger2008} combined the frequencies obtained from spectroscopic and photometric observations and interpreted them as long lived (order tens of days) radial and non-radial modes. These results were in agreement with \citet{hekker2006b}, but not with \citet{deridder2006} and \citet{barban2007}.

From the latter example it is clear that the interpretation of the mode lifetime and the degree of the mode are not independent. On the one hand an interpretation of short mode lifetimes (wide frequency peaks in the power spectrum) might incorporate a possible $\ell$~=~2 mode as part of the $\ell$~=~0 mode. On the other hand an interpretation of long lifetimes (narrow peaks in the power spectrum) might lead to the interpretation of peaks as $\ell$~=~2 modes, which are in reality part of the neighbouring $\ell$~=~0 mode.

\begin{figure}
\begin{minipage}{\linewidth}
\centering
\includegraphics[width=\linewidth]{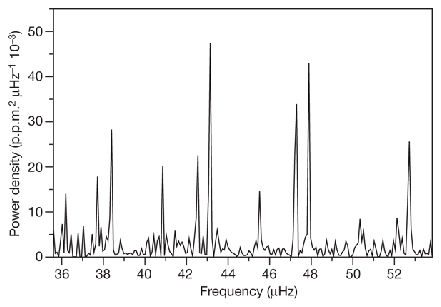}
\end{minipage}
\hfill
\begin{minipage}{\linewidth}
\centering
\includegraphics[width=\linewidth]{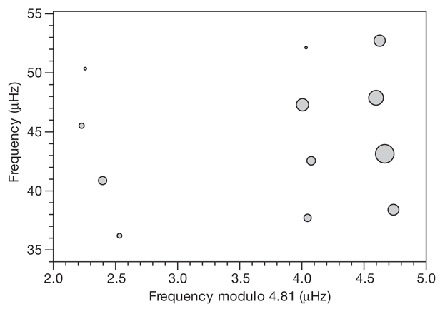}
\end{minipage}
\caption{Power spectrum (top) and \'{e}chelle diagram of a red giant observed with CoRoT, which shows evidence of non-radial modes. (Figure taken from \citet{deridder2009}).}
\label{deridder}
\end{figure}

With recent CoRoT and Kepler observations (Bedding et al. 2010, De Ridder et al. 2009) the degree of the oscillation modes of many red giants could be investigated through the asymptotic relation (Eq.~\ref{asymptot} and Fig.~\ref{deridder}). These results show unambiguous evidence for the presence of non-radial oscillations in red giants. Also more recent theoretical work by \citet{dupret2009} predicts the presence of non-radial oscillations in red giants with observable heights in the power spectrum, i.e., these are non-radial modes trapped in the envelope.

\section{Current challenges}
The field of red-giant asteroseismology has not yet matured, and is both observationally as well as theoretically a challenging field. To infer the internal structure of the stars a number of parameters need to be determined, such as the frequency of maximum oscillation power and large frequency separation ($\Delta \nu$ in Eq.~\ref{asymptot}), which provides information on the stellar radius \citep{kjeldsen1995}, which increases with age, the oscillation frequencies, small frequency spacing and mode identification. When knowing the individual frequencies, their degrees and stellar parameters such as effective temperature, surface gravity and metallicity, an exact model for the star can in principle be computed. So far, both theoretical and observational problems have prevented the determination of the internal structure of red giants in detail.

\subsection{Theoretical challenges}
One of the main challenges from the theoretical side is the computation of convection. 
The often used classical mixing-length parametrisation \citep{bohmvitense1958} is computationally attractive, but does not provide a physical description of the convection processes within a star. A further theoretical dilemma is that the numerical stability of some evolution codes decreases
when modelling a helium flash scenario at the tip of the giant branch. This can be solved by computationally expensive algorithms or by circumventing the numerical problems by restarting the computation at the horizontal branch (see e.g. \citet{mazumdar2009}). This has the disadvantage that the internal structure changes taking place during the He-flash cannot be followed.

Another numerical difficulty originates from the very large number of nodes in the eigenfunctions in the g-mode cavity. The huge number of nodes originates from the high density contrast between the core and the envelope and leads to radiative damping. Due to the mixed nature of the non-radial modes, the amount and exact region in the star where this damping occurs may have influence on the observed oscillations. Therefore the inner region of the star needs to be probed with a dense numerical grid, which is computationally challenging. See \citet{dupret2009} and Montalban et al. (2010) (these proceedings) for state-of-the-art models of oscillating red-giant stars.

\subsection{Observational challenges}
First of all, the finite mode lifetimes of stochastically driven and damped oscillations are predicted to be different for modes of different degrees, depending on their mode inertia. 
Higher mode inertia for non-radial modes lead to longer lifetimes, generally of the order of tens of days, but they can be of the order of hundreds of days or more (see e.g. \citet{dupret2009}). 
If the duration of the observations is shorter than these lifetimes, the modes are unresolved and have lower heights in the power spectrum than resolved modes. In order to resolve, and thus observe, non-radial modes, time series should be used with a length that exceeds the expected lifetime of the observed star. To be able to obtain a reliable measure of the lifetime, the timespan of the timeseries should be of the order of ten times as long as the lifetimes of the oscillation modes \citep{hekker2010c}.
The relatively short timespan of about a month for the single-site observations of $\xi$ Hydrae are most likely the reason for the different results regarding the mode lifetimes (see Section 3). 

A second observational challenge relates to the trapping of the non-radial modes. The size of the evanescent region and thus the efficiency of trapping increases with the degree $\ell$ of the modes. When only modes trapped in the envelope are considered, the frequency spectrum shows a regular pattern similar to solar-type main-sequence stars which can be interpreted using Eq.~\ref{asymptot}. However, for stars at a certain evolutionary phase, such as the bottom of the giant branch, the trapping is not efficient and the interaction between the p- and g-mode cavities leads to many avoided crossings. Hence several modes with $\ell$~=~1 with nearly the same frequency but different $n$ can be observable \citep{dupret2009}. This leads to a complex frequency pattern in which there might exist ambiguity between the interpretation of an oscillation feature as a single wide mode with short mode lifetime or several mixed modes with the same degree and longer lifetimes. 

Thirdly, the large radii of red-giant stars adjust the pulsation periods from minutes, for the Sun, to hours. This complicates ground-based single-site efforts considerably, and calls for either multi-site campaigns or observations from space. Furthermore, like for the Sun, turbulent motions in the convective envelope cause granulation. The granulation time scales depend on the size of the granules, and their velocities, and are generally of the order of days for red giants. For the most evolved and luminous stars, which have the largest radii, the granulation time scales are of the same order as the oscillation timescales. This causes difficulties in disentangling granulation and oscillation features in the power spectrum. Increase in frequency resolution, i.e. timeseries with longer timespan, are needed to investigate the features at low frequencies in more detail.

The realisation of the CoRoT satellite, and more recently the Kepler satellite, are very important for tackling the observational challenges of problems related to observing solar-like oscillations in red giants. These satellites observe the same field  with high-precision photometry for 150 days and 3.5 years, respectively. This means that non-radial modes can be resolved in which case they have detectable heights in the power spectrum. The increased frequency resolution will also be important to attempt to disentangle granulation and oscillation features in the power spectrum.

\section{Future prospects}
So far the CoRoT data have already greatly increased our knowledge of solar-like oscillations in red-giant stars, both from an observational as well as from a theoretical point of view. Apart from the ensemble studies \citep{hekker2009,mosser2010} and the unambiguous determination of long lived non-radial oscillations \citep{deridder2009}, detailed studies of individual stars, an investigation into mass and radius determination from oscillation properties (Kallin- ger et al. 2010) and a statistical investigation in mode lifetimes (Baudin et al. 2010) have been carried out. These observational efforts have been accompanied by theoretical investigations  using adiabatic (e.g., \citet{dupret2009}) and non-adiabatic codes to simulate the stars, as well as stellar population studies \citep{miglio2009}. Also, first results from the Kepler satellite on low-mass low-luminosity red giants \citep{bedding2010} enabled for an extended statistical study in a frequency range where CoRoT is less sensitive.

Notwithstanding these efforts, there still remain many open questions about the internal structures of red giants, and the different processes taking place in different layers and between layers in these stars. The excitation and damping of these oscillations, the time scales at which these process occur and their connection with convection and granulation are not well understood. Furthermore, it would be very interesting to be able to infer whether the star is in the hydrogen-shell burning or helium-burning phase. With a detailed picture of the internal structure and mass of the red giants, it might also be possible to trace back their evolution and determine their primordial rotation velocity or infer the difference in internal structure due to different chemical composition.\newline
\newline
The future looks very promising with respect to observations from dedicated instruments for asteroseismology. The CoRoT satellite is still taking data and will do so for the next couple of years and Kepler has now been taking data for nearly a year, with a 3.5 year nominal duration of the mission. Both satellites observe many relatively faint stars (roughly 9 - 16 mag in V), which has the advantage of providing a statistically significant sample, but the disadvantage that stellar parameters, such as effective temperature, surface gravity and metallicity can not be obtained with high accuracy.

For bright targets, for which it is relatively easy to obtain stellar parameters, the MOST satellite (Micro variability and Oscillations of STars) has been taking valuable time series data of a couple of weeks for several red giants. Furthermore, dedicated instrumentation is currently under development.  BRITE-Constellation consists of at least two small satellites with different colour filters, which aim to take simultaneous two-colour photometry of bright targets (2-6 mag in V). Depending on the exact launch time, which is scheduled for 2011, there are fields that can be observed year round, while other fields can be observed for at least half a year. Also PLATO (PLAnetary Transits and Oscillations of stars), a medium-sized mission, is in preparation and among three missions of which two will be launched in the ESA Cosmic Vision 2015 - 2025 programme.

Another development is SONG (Stellar Observations Network Group), which aims for a dedicated network of 1-m class telescopes with high-resolution spectrographs for radial-velocity measurements. The prototype of these telescopes is currently being build at Tenerife, Spain.

Progress is also being made on the theoretical interpretation of the solar-like oscillations in red giants. Stellar evolution codes are constantly improved and many more are now extended to also incorporate evolution of giant stars. Detailed modelling of red giants for which oscillations are now observed with high accuracy (for example HR 7349 \citet{carrier2010}) are now carried out. Also ensemble studies such at the one by \citet{miglio2009} are developed further.\newline
\newline
It is clear that with the increased frequency resolution and accuracy of the data from recent and future instruments and the simultaneous developments in theory, significant progress in the field of red-giant asteroseismology has recently been made and can be expected in the (near) future. 

\acknowledgements
SH wants to thank Maarten Mooij and Bill Chaplin for useful discussions that improved the manuscript considerably. SH acknowledges financial support from the UK Science and Technology Facilities Council (STFC).

\bibliographystyle{aa}
\bibliography{bibinvited}

\begin{thebibliography}{32}
\expandafter\ifx\csname natexlab\endcsname\relax\def\natexlab#1{#1}\fi

\bibitem[{{Barban} {et~al.}(2004){Barban}, {De Ridder}, {Mazumdar}, {Carrier},
  {Eggenberger}, {De Ruyter}, {Vanautgaerden}, {Bouchy}, \&
  {Aerts}}]{barban2004}
{Barban}, C., {De Ridder}, J., {Mazumdar}, A., {et~al.} 2004, in ESA Special
  Publication, Vol. 559, SOHO 14 Helio- and Asteroseismology: Towards a Golden
  Future, ed. {D.~Danesy}, 113

\bibitem[{{Barban} {et~al.}(2007){Barban}, {Matthews}, {De Ridder}, {Baudin},
  {Kuschnig}, {Mazumdar}, {Samadi}, {Guenther}, {Moffat}, {Rucinski},
  {Sasselov}, {Walker}, \& {Weiss}}]{barban2007}
{Barban}, C., {Matthews}, J.~M., {De Ridder}, J., {et~al.} 2007, A\&A, 468,
  1033

\bibitem[{{Bedding} {et~al.}(2010){Bedding}, {Huber}, {Stello}, {Elsworth},
  {Hekker}, {Kallinger}, {Mathur}, {Mosser}, {Preston}, {Ballot}, {Barban},
  {Broomhall}, {Buzasi}, {Chaplin}, {Garc{\'{\i}}a}, {Gruberbauer}, {Hale}, {De
  Ridder}, {Frandsen}, {Borucki}, {Brown}, {Christensen-Dalsgaard},
  {Gilliland}, {Jenkins}, {Kjeldsen}, {Koch}, {Belkacem}, {Bildsten}, {Bruntt},
  {Campante}, {Deheuvels}, {Derekas}, {Dupret}, {Goupil}, {Hatzes}, {Houdek},
  {Ireland}, {Jiang}, {Karoff}, {Kiss}, {Lebreton}, {Miglio}, {Montalb{\'a}n},
  {Noels}, {Roxburgh}, {Sangaralingam}, {Stevens}, {Suran}, {Tarrant}, \&
  {Weiss}}]{bedding2010}
{Bedding}, T.~R., {Huber}, D., {Stello}, D., {et~al.} 2010, ApJ, 713, L176

\bibitem[{{B{\"o}hm-Vitense}(1958)}]{bohmvitense1958}
{B{\"o}hm-Vitense}, E. 1958, Zeitschrift fur Astrophysik, 46, 108

\bibitem[{{Borucki} {et~al.}(2009){Borucki}, {Koch}, {Batalha}, {Caldwell},
  {Christensen-Dalsgaard}, {Cochran}, {Dunham}, {Gautier}, {Geary},
  {Gilliland}, {Jenkins}, {Kjeldsen}, {Lissauer}, \& {Rowe}}]{borucki2009}
{Borucki}, W., {Koch}, D., {Batalha}, N., {et~al.} 2009, in IAU Symposium, Vol.
  253, IAU Symposium, 289--299

\bibitem[{{Brewer} \& {Stello}(2009)}]{brewer2009}
{Brewer}, B.~J. \& {Stello}, D. 2009, \mnras, 395, 2226

\bibitem[{{Buzasi} {et~al.}(2000){Buzasi}, {Catanzarite}, {Laher}, {Conrow},
  {Shupe}, {Gautier}, {Kreidl}, \& {Everett}}]{buzasi2000}
{Buzasi}, D., {Catanzarite}, J., {Laher}, R., {et~al.} 2000, ApJL, 532, L133

\bibitem[{{Carrier} {et~al.}(2010){Carrier}, {De Ridder}, {Baudin}, {Barban},
  {Hatzes}, {Hekker}, {Kallinger}, {Miglio}, {Montalb{\'a}n}, {Morel}, {Weiss},
  {Auvergne}, {Baglin}, {Catala}, {Michel}, \& {Samadi}}]{carrier2010}
{Carrier}, F., {De Ridder}, J., {Baudin}, F., {et~al.} 2010, A\&A, 509, A73

\bibitem[{{De Ridder} {et~al.}(2009){De Ridder}, {Barban}, {Baudin}, {Carrier},
  {Hatzes}, {Hekker}, {Kallinger}, {Weiss}, {Baglin}, {Auvergne}, {Samadi},
  {Barge}, \& {Deleuil}}]{deridder2009}
{De Ridder}, J., {Barban}, C., {Baudin}, F., {et~al.} 2009, Nature, 459, 398

\bibitem[{{De Ridder} {et~al.}(2006){De Ridder}, {Barban}, {Carrier},
  {Mazumdar}, {Eggenberger}, {Aerts}, {Deruyter}, \&
  {Vanautgaerden}}]{deridder2006}
{De Ridder}, J., {Barban}, C., {Carrier}, F., {et~al.} 2006, A\&A, 448, 689

\bibitem[{{Dupret} {et~al.}(2009){Dupret}, {Belkacem}, {Samadi}, {Montalban},
  {Moreira}, {Miglio}, {Godart}, {Ventura}, {Ludwig}, {Grigahc{\`e}ne},
  {Goupil}, {Noels}, \& {Caffau}}]{dupret2009}
{Dupret}, M., {Belkacem}, K., {Samadi}, R., {et~al.} 2009, A\&A, 506, 57

\bibitem[{{Dziembowski} {et~al.}(2001){Dziembowski}, {Gough}, {Houdek}, \&
  {Sienkiewicz}}]{dziembowski2001}
{Dziembowski}, W.~A., {Gough}, D.~O., {Houdek}, G., \& {Sienkiewicz}, R. 2001,
  MNRAS, 328, 601

\bibitem[{{Frandsen} {et~al.}(2002){Frandsen}, {Carrier}, {Aerts}, {Stello},
  {Maas}, {Burnet}, {Bruntt}, {Teixeira}, {de Medeiros}, {Bouchy}, {Kjeldsen},
  {Pijpers}, \& {Christensen-Dalsgaard}}]{frandsen2002}
{Frandsen}, S., {Carrier}, F., {Aerts}, C., {et~al.} 2002, A\&A, 394, L5

\bibitem[{{Frink} {et~al.}(2001){Frink}, {Quirrenbach}, {Fischer}, {R{\"o}ser},
  \& {Schilbach}}]{frink2001}
{Frink}, S., {Quirrenbach}, A., {Fischer}, D., {R{\"o}ser}, S., \& {Schilbach},
  E. 2001, PASP, 113, 173

\bibitem[{{Hekker} \& {Aerts}(2010)}]{hekker2010b}
{Hekker}, S. \& {Aerts}, C. 2010, A\&A, 515, A43

\bibitem[{{Hekker} {et~al.}(2006{\natexlab{a}}){Hekker}, {Aerts}, {De Ridder},
  \& {Carrier}}]{hekker2006b}
{Hekker}, S., {Aerts}, C., {De Ridder}, J., \& {Carrier}, F.
  2006{\natexlab{a}}, A\&A, 458, 931

\bibitem[{{Hekker} {et~al.}(2010{\natexlab{a}}){Hekker}, {Barban}, {Baudin},
  J., {Chaplin}, \& {Elsworth}}]{hekker2010c}
{Hekker}, S., {Barban}, C., {Baudin}, F., {et~al.} 2010{\natexlab{a}}, A\&A, in
  press

\bibitem[{{Hekker} {et~al.}(2010{\natexlab{b}}){Hekker}, {Debosscher}, {Huber},
  {Hidas}, {De Ridder}, {Aerts}, {Stello}, {Bedding}, {Gilliland},
  {Christensen-Dalsgaard}, {Brown}, {Kjeldsen}, {Borucki}, {Koch}, {Jenkins},
  {Van Winckel}, {Beck}, {Blomme}, {Southworth}, {Pigulski}, {Chaplin},
  {Elsworth}, {Stevens}, {Dreizler}, {Kurtz}, {Maceroni}, {Cardini}, {Derekas},
  \& {Suran}}]{hekker2010a}
{Hekker}, S., {Debosscher}, J., {Huber}, D., {et~al.} 2010{\natexlab{b}}, ApJ,
  713, L187

\bibitem[{{Hekker} {et~al.}(2009){Hekker}, {Kallinger}, {Baudin}, {De Ridder},
  {Barban}, {Carrier}, {Hatzes}, {Weiss}, \& {Baglin}}]{hekker2009}
{Hekker}, S., {Kallinger}, T., {Baudin}, F., {et~al.} 2009, A\&A, 506, 465

\bibitem[{{Hekker} {et~al.}(2006{\natexlab{b}}){Hekker}, {Reffert},
  {Quirrenbach}, {Mitchell}, {Fischer}, {Marcy}, \& {Butler}}]{hekker2006a}
{Hekker}, S., {Reffert}, S., {Quirrenbach}, A., {et~al.} 2006{\natexlab{b}},
  A\&A, 454, 943

\bibitem[{{Houdek} \& {Gough}(2002)}]{houdek2002}
{Houdek}, G. \& {Gough}, D.~O. 2002, MNRAS, 336, L65

\bibitem[{{Kallinger} {et~al.}(2008){Kallinger}, {Guenther}, {Weiss},
  {Hareter}, {Matthews}, {Kuschnig}, {Reegen}, {Walker}, {Rucinski}, {Moffat},
  \& {Sasselov}}]{kallinger2008}
{Kallinger}, T., {Guenther}, D.~B., {Weiss}, W.~W., {et~al.} 2008,
  Communications in Asteroseismology, 153, 84

\bibitem[{{Kjeldsen} \& {Bedding}(1995)}]{kjeldsen1995}
{Kjeldsen}, H. \& {Bedding}, T.~R. 1995, A\&A, 293, 87

\bibitem[{{Matthews} {et~al.}(2000){Matthews}, {Kuschnig}, {Walker}, {Pazder},
  {Johnson}, {Skaret}, {Shkolnik}, {Lanting}, {Morgan}, \&
  {Sidhu}}]{matthews2000}
{Matthews}, J.~M., {Kuschnig}, R., {Walker}, G.~A.~H., {et~al.} 2000, in
  Astronomical Society of the Pacific Conference Series, Vol. 203, IAU Colloq.
  176: The Impact of Large-Scale Surveys on Pulsating Star Research, ed.
  L.~{Szabados} \& D.~{Kurtz}, 74--75

\bibitem[{{Mazumdar} {et~al.}(2009){Mazumdar}, {M{\'e}rand}, {Demarque},
  {Kervella}, {Barban}, {Baudin}, {Coud{\'e} du Foresto}, {Farrington},
  {Goldfinger}, {Goupil}, {Josselin}, {Kuschnig}, {McAlister}, {Matthews},
  {Ridgway}, {Sturmann}, {Sturmann}, {ten Brummelaar}, \&
  {Turner}}]{mazumdar2009}
{Mazumdar}, A., {M{\'e}rand}, A., {Demarque}, P., {et~al.} 2009, A\&A, 503, 521

\bibitem[{{Merline}(1999)}]{merline1999}
{Merline}, W.~J. 1999, in Astronomical Society of the Pacific Conference
  Series, Vol. 185, IAU Colloq. 170: Precise Stellar Radial Velocities, ed.
  {J.~B.~Hearnshaw \& C.~D.~Scarfe}, 187

\bibitem[{{Miglio} {et~al.}(2009){Miglio}, {Montalb{\'a}n}, {Baudin},
  {Eggenberger}, {Noels}, {Hekker}, {De Ridder}, {Weiss}, \&
  {Baglin}}]{miglio2009}
{Miglio}, A., {Montalb{\'a}n}, J., {Baudin}, F., {et~al.} 2009, A\&A, 503, L21

\bibitem[{{Mosser} {et~al.}(2010){Mosser}, {Belkacem}, {Goupil}, {Miglio},
  {Morel}, {Barban}, F., {Hekker}, {Samadi}, {De Ridder}, {Weiss}, {Auvergne},
  \& {Baglin}}]{mosser2010}
{Mosser}, B., {Belkacem}, K., {Goupil}, M.-J., {et~al.} 2010, A\&A, in press

\bibitem[{{Stello} {et~al.}(2006){Stello}, {Kjeldsen}, {Bedding}, \&
  {Buzasi}}]{stello2006}
{Stello}, D., {Kjeldsen}, H., {Bedding}, T.~R., \& {Buzasi}, D. 2006, A\&A,
  448, 709

\bibitem[{{Tarrant} {et~al.}(2007){Tarrant}, {Chaplin}, {Elsworth},
  {Spreckley}, \& {Stevens}}]{tarrant2007}
{Tarrant}, N.~J., {Chaplin}, W.~J., {Elsworth}, Y., {Spreckley}, S.~A., \&
  {Stevens}, I.~R. 2007, MNRAS, 382, L48

\bibitem[{{Tassoul}(1980)}]{tassoul1980}
{Tassoul}, M. 1980, ApJS, 43, 469

\bibitem[{{Teixeira} {et~al.}(2003){Teixeira}, {Christensen-Dalsgaard},
  {Carrier}, {Aerts}, {Frandsen}, {Stello}, {Maas}, {Burnet}, {Bruntt}, {de
  Medeiros}, {Bouchy}, {Kjeldsen}, \& {Pijpers}}]{teixeira2003}
{Teixeira}, T.~C., {Christensen-Dalsgaard}, J., {Carrier}, F., {et~al.} 2003,
  Ap\&SS, 284, 233

\end{thebibliography}
\end{document}